\documentstyle[12pt]{article}

\begin{document}

\title{A Phase Space Approach to Gravitational Entropy}

\author{Tony Rothman \\ Dept. of Physics, Illinois Wesleyan
University}
\maketitle
\begin{abstract}
We  examine the definition $S = \ln\Omega$ as a candidate function
for ``gravitational entropy."  We calculate its behavior for
gravitational and density perturbations in closed, open and flat
cosmological models and find that in all cases it increases
monotonically.  We are also able to calculate the entropy density
of gravitational radiation produced by inflation.  We compare the
results with the behavior of the Weyl-tensor squared.  Applying
the formalism to black holes has proved more problematical.
\end{abstract}

 It is appropriate that I speak today on the
subject of gravitational entropy because it was George Ellis, on
my last visit to Cape Town, who instigated my investigations into
this area---though he shouldn't be held responsible for any of the
results.

The problem is well-known.  Ordinary thermodynamic systems, a box
of gas for example, tend to grow more homogeneous with time,
whereas gravitating systems tend to become more inhomogeneous with
time. In this sense gravitating systems are ``antithermodynamic.''
I'm sure you've all seen the picture in Roger Penrose's book that
shows this behavior.  Indeed, the tendency toward inhomogeneity of
a gravitating system can be viewed as a manifestation of the
long-range nature of the gravitational force, which tends to cause
the components of the system to clump.

Now, ordinarily, for thermodynamic system, we associate the
increase of homogeneity with an increase in thermodynamic entropy.
This is a sign convention.  One could, after all, choose
Boltzmann's original H as the entropy function, and it would
decrease with homogeneity. Whichever sign one chooses,
though,
``gravitational entropy" or the ``gravitational arrow of
time" points in the opposite direction to the thermodynamic arrow.

The question then becomes: Can you find a generally covariant
function that characterizes the tendency of the gravitational
field to become more inhomogeneous with time, a ``gravitational
entropy"?

However, in tackling this problem one immediately encounters a
major conceptual difficulty: general relativity is a dynamical,
not a thermodynamical theory, and it does not deal with
temperatures.  As prima facie evidence, I have been keeping a
rough count of how many times the word ``temperature" has arisen
in the past four days and I would say five, $\pm 1$, and in all
cases it has referred to the cosmic microwave background.  Another
way of stating the difficulty is that general relativity deals
with far-from-equilibrium systems for which a temperature is not
well-defined.  One usually, for example, doesn't talk about the
temperature of a pendulum.  Of course, you could define an
effective temperature $1/2mv^2 \sim kT$ but the fact remains--

Jurgen Ehlers: The pendulum is not in equilibrium.

TR: Exactly.

So the question remains whether the concept of entropy can be
incorporated into GR and how to do it.  There don't seem to have
been too many attempts in the literature to define a gravitational
entropy.  The most well-known is Roger Penrose's Weyl-tensor
hypothesis \cite{Penrose89}, the idea being that the square of the
Weyl-tensor should be zero at the Big Bang but increase
monotonically thereafter. However, the work of Bonnor
\cite{Bonnor87}, as well as that of Wainright and collaborators
\cite{Wainwright84,Goode85,Goode92}, have thrown some doubt on the
proposal.  It's not my intent to criticize the hypothesis but
rather to rethink the problem. Smolin \cite{Smolin85} and Hu and
Kandrup \cite{Hu87} have examined some aspects of gravitational
entropy, but not the arrow-of-time question.  After I began work
on this topic a paper by Brandenburger et al. appeared
\cite{Brandenberger93}, which again is concerned with the
thermodynamical aspects of the entropy of gravitational radiation,
but in the places where we can compare our work with their's, it
seems to agree.  Most of this talk will be based on a paper I
wrote with Peter Anninos that appeared in Phys Rev D \cite{RA97}.

Roger Penrose: By way of clarification I should say that I never
meant the Weyl tensor to be a measure of gravitational entropy. I
merely wanted it to be zero at the big bang.

TR:  Really? Well, if that's true I apologize.

Roger Penrose: You're not the only person to have that
misunderstanding.

TR: It does seem to be widespread.  Perhaps we can sort out where
people got this idea.

At any rate, I decided to tackle the problem by a direct
statistical-mechanics approach. We examine the definition
\begin{equation}
S = \ln \Omega \label{S_def}
\end{equation}
where $S = $ gravitational entropy and $\Omega = $ volume of phase
space for the field.

Why choose such a definition?  In thermodynamic systems the phase
space is much too hard to calculate.  Instead you generally
evaluate the partition function $Z \equiv \sum_i e^{-\beta E_i}$,
from which the entropy is readily derived(here $\beta \equiv
1/(T)$ and ${\overline E} =$ the mean energy.) But this requires a
temperature, which we don't have.

On the other hand, consider a collection of $N$ harmonic
oscillators with the usual Hamiltonian
\begin{equation}
H = \frac{1}{2} \sum_{i = 1}^{N} \dot \phi_i^2 +
\frac{k}{2}\sum_{i=1}^N \phi _i^2\label{2p:ham}
\end{equation}
This has a phase space.  In fact, for one pendulum you know the
phase space: an ellipse.  If you increased the energy of the
pendulum it would describe a slightly larger ellipse and the
logarithm of the area between the two ellipses could be taken as
entropy. However,we  don't talk about entropy of dynamical
systems--you know exactly the trajectory of the pendulum and so it
doesn't make sense to define  the logarithm of the area as
entropy. You need an
 ergodic system, some lack of information.

However, if we consider a system of oscillators with random
phases, then we don't know where the oscillators are in this
2N-dimensional phase space and it apparently makes sense to
interpret the logarithm of this volume as entropy.  In essence I
am going to model the field as a collection of oscillators and
treat the oscillators as a microcanonical ensemble, in which you
assume any region of phase space is occupied with equal
probability. In that case $S = \ln\Omega$ is equivalent to $S =
-\sum_i p_i \ln p_i$.

To give you an idea of how it works, for the Hamltonian given in
(\ref{2p:ham}) above, we can actually calculate the phase-space
volume analytically, using beta-functions, much in the way you'd
calculate the volume of a sphere in elementary calculus.  The
result is:
\begin{equation}
\Omega = \frac{(2\pi)^N H^N}{k^{N/2}N!}  \label{Omega_ho}
\end{equation}

Noting that $H = R^2$, this indeed is the volume of a
2N-dimensional sphere, or more precisely an ellipsoid.  Taking $S
= ln\Omega \approx N$ one can recover from this formula in the
appropriate limits :

$\bullet$ Entropy of an ideal gas($V = k = 0$)

$\bullet$ Classical limit of entropy of an Einstein solid ($H = E
= NT$)

This isn't too surprising, as Einstein modeled a solid as a
collection of harmonic oscillators. You can also get approximately

$\bullet$ Entropy of the electromagnetic field.

In fact you can write down

$\bullet$ Partition function ($\frac{2\pi}{\beta\omega}$)

although I haven't figured out what to do with it.

You might well be asking what this has to do with the
gravitational field.  The first reason to examine the above
definition of entropy is that it does not require a temperature.
The second reason is that there exists a Hamiltonian formulation
of gravity (the ADM formulation). One can write down Hamiltonians
for gravitational systems and thus one should be able to define
$\Omega$ below $H$.

Here one needs to make an important caveat: In general, $H$ will
be time dependent, and therefore not the energy. So this is a
slight extension of the usual statistical mechanics definition of
energy. This doesn't necessarily bother me.  In stat mech you for
a system in contact with a heat reservoir the change in entropy is
$\Delta S = \Delta Q/T$.  If you believe that $Q$ is the kinetic
energy of the Hamiltonian of the system, this does assume a change
in $H$, although a quasi-static one. A time-dependent $H$ is
characteristic of an open system, which is really what we have in
GR, and it is this time-dependence that will cause a change in
entropy. So how do we calculate the change in $H$?

Well, finally, I'll write down a metric, one for
gravitational-wave perturbations:

\begin{equation}
ds^2 = a^2(\eta)\Bigl[-d\eta^2 +
       \Bigl(\delta_{ij} + h_{ij}(\eta,z)\Bigr) dx^idx^j\Bigr] ,
\label{metric_gw}
\end{equation}
Here $\eta =$  conformal time; $a(\eta)=$ expansion scale factor;

and $h_{ij} \ll \delta_{ij}=$ represent the gravitational wave
perturbations.

Assuming a single polarization, the Einstein-Hilbert action to
second order in the perturbation variables is:
\begin{equation}
I = \frac{1}{64\pi} \int a^2 (\dot h^2 - h'^2) d^4 x ,
\label{gwaves_action}
\end{equation}
Here $(\cdot) \equiv d/d\eta$ and $(') \equiv d/dz$.

It's sometimes useful, in particular when comparing with other
work, to change variables to $\phi \equiv ah/\sqrt{32\pi}$, and
you get for the above  Lagrangian density:

\begin{equation}
  {\cal L} = \frac{1}{2}\Bigl[\dot\phi^2 - \phi'^2 + \frac{\ddot
a}{a}\phi^2\Bigr]. \label{L_h}
\end{equation}

The corresponding Hamiltonian density is by definition ${\cal
H}\equiv \pi\dot q - {\cal L}=$, or in this case,
\begin{equation}
  {\cal H} = \frac{1}{2}\Bigl[\dot\phi^2 + \phi'^2 - \frac{\ddot
a}{a}\phi^2\Bigr] .
\end{equation}
The Hamiltonian is $H = \int{\cal H} d^3x$.  In our case we want
to make contact with the system of N harmonic oscillators and so
we discretize the integral:
\begin{equation}
{\overline H} =
 \frac{1}{2}\sum_{i=1}^N\Bigl(\overline\pi_i^2 + \overline\phi_i'^2
               - \frac{\ddot a}{a}\overline\phi_i^2\Bigr) ,
\label{H_grav}
\end{equation}
where $\overline\pi = \pi(L/N)^{3/2}$, $\overline\phi =
\phi(L/N)^{3/2}$, $\overline H = H/N^2$, and L is an arbitrary
length scale.

Notice that this equation resembles the harmonic oscillator
Hamiltonian Eq (\ref{2p:ham}) except for two features.  The first
is that it contains  a gradient term: $\phi' = \phi_i -
\phi_{(i+1)}$. This is not a big problem. By defining a new
variable $ \xi_i
 \equiv \phi_i - \phi_{(i+1)}$, one can integrate this
``nearest-neighbor potential" in the same way one does the
 harmonic oscillator potential and get an result for the phase
 space  identical to Eq (\ref{Omega_ho}), but for insignificant  numerical factors.

The second feature is the main problem: the minus sign in front of
the harmonic-like potential of the last term.  This implies,
first, that $H$ can change sign and that, second, we have a
reflection barrier instead of a potential well.  This implies that
the
 phase space is unbounded.  Without going into the gory details, I claim
that in the perturbative limit, one can still analytically
integrate the phase space using hypergeometric functions (see
Rothman and Anninos for specifics), and one formally recovers
expression (\ref{Omega_ho}).  Nevertheless, one is forced to
impose some sort of cutoff, in momentum space if $H > 0$ or in
configurations space if $H < 0$.  (It turns out that $H > 0$
represents growing modes and $H < 0$ represents decaying modes,
which is an interesting result in its own right, but we do not go
into details here.  The growing modes, which result in an increase
in entropy, are of the main concern.) There are several
conceivable methods to impose the necessary cutoffs, but it turns
out they all give qualitatively the same results.

So, the procedure to calculate the entropy is to regard $\Omega$
as constant on each spacelike hypersurface, in which case we can
take  $\ddot a/a = k$, the spring constant.  To find $\Omega$, we
first need $H$ , which we get by considering the equations of
motion resulting from the action (\ref{gwaves_action}):
\begin{equation}
  \ddot h + 2\frac{\dot a}{a} \dot h - h'' = 0 .
\label{gwaves_bess}
\end{equation}

We assume a separable solution with random phases $\alpha_j$:
\begin{equation}
\sum_j \Bigl(\ddot h_j + \frac{2\dot a}{a} \dot h_j - h_j''\Bigr)
 = \Bigl(\ddot h + \frac{2\dot a}{a} \dot h -
 h''\Bigr) \sum_j e^{i\alpha_j} = 0 ,
\end{equation}
The subscript $j$ refers to different $\sl waves$, not to
different coordinates.  By taking the phases to be random we are
essentially assuming an incoherent source, and since the
$\alpha_j$ do not enter into the solutions, I supress them from
now on.  Eq (\ref{gwaves_bess}) is Bessel's equation and for the
matter-dominated epoch with  $a = a_o\eta^2$, one gets:
\begin{equation}
  h \propto \eta^{-3/2} J_{\pm 3/2}(k\eta)e^{ikz} ,
\label{gw_bessel}
\end{equation}

Now, the $J$'s have standard asymptotic forms for $k\eta \ll 1$
($\lambda \gg$ Hubble radius )and $k\eta \gg 1$ ($\lambda \ll$
Hubble radius).

For  $k\eta \ll 1$, we have:
\begin{equation}
  h = \Bigl[ h_1 (k\eta)^{-3} + h_2 \Bigr] e^{ikz} ,
\label{h_super}
\end{equation}
where $h_1$ and $h_2 $ are constants representing the decaying and
growing modes respectively.  (In this case the ``growing" modes
are constant in time.)

For $k\eta \gg 1$ we have:
\begin{eqnarray}
  h &\propto \sqrt{\frac{2k^3}{\pi}}\left(\frac{1}{k\eta}\right)^2
      \Bigl[\cos(k\eta) + \sin(k\eta)\Bigr] e^{ikz} \nonumber \\
&\propto (k\eta)^{-2} \times [{oscillations}]
\end{eqnarray}

In terms of $\phi \sim ah$, for $k\eta \gg 1$:
 \begin{eqnarray}
& H  \propto \pi^2 + k^2\phi^2 \nonumber \\
 & = \mbox{Harmonic oscillator potential at fixed time},
 \end{eqnarray}

and
\begin{equation}
\Omega \propto \frac{H^N}{k^{N/2}}
       \propto \mbox{constant} \times [\mbox{oscillations}] .
\label{Omega_gw_sub}
\end{equation}

Therefore $\Omega$ oscillates, but otherwise does not change.
Given that in this limit $H$ is a harmonic oscillator, this isn't
too surprising. One could average over several cycles to obtain a
strictly constant $H$:
\begin{equation}
 ^{(4)}{\cal H} =
 \int d^4x\Bigl[\pi^2 + k^2\phi^2 - \frac{\ddot a}{a}\phi^2\Bigr] .
\end{equation}

In the nonlinear regime, $\Omega$ would increase, which is
encouraging for the interpretation of $\ln\Omega$ as entropy.

In the opposite limit $k\eta \ll 1$, the gradient term in
(\ref{H_grav}) is negligible  and we get for the time dependence
of $H$:
 $H \propto \pi^2 - \ddot a \phi^2 / a$ and $\Omega \propto
H^N / (\ddot a/a)^{N/2}$.
\begin{eqnarray}
H \propto \left\{
  \begin{array}{ll}
        \eta^2    &,    \\
        \eta^{-4} &,
  \end{array}
  \right.
\mbox{and}\  \Omega \propto \left\{
  \begin{array}{lll}
        \eta^{3N}  &,     &  \mbox{for growing modes,} \\
        \eta^{-3N} &,     &  \mbox{for decaying modes.}
  \end{array}
  \right.
\end{eqnarray}

This result is at first glance somewhat perplexing. Superhorizon
modes should be frozen in, no oscillations allowed, so why should
we get an increase in $\Omega$? In terms of $h$ one sees that
indeed: $\dot\phi = \dot a h + a \dot h = \dot a h$. That is since
$\dot h = 0$, the superhorizon modes {\sl are} frozen in and the
increase in $\Omega$ is due entirely to expansion $\dot a$.
Furthermore, since the superhorizon modes are frozen, we know the
position of each oscillator and the assumption of random phases
breaks down. This suggests that our definition of entropy is only
applicable to subhorizon processes, which is not necessarily a bad
thing. Such an interpretation agrees with the conclusions of
Brandenburger el al.

Without going into further details, I merely state that one can
repeat the above analysis for density perturbations (radiation and
dust) in flat, open and closed cosmologies.  One always has to
worry about gauge problems in inhomogeneous models and so we use
the gauge-invariant formalism of Mukhanov et al.
\cite{Mukhanov92}. We find that {\sl in all cases $\Omega$
monotonically increases for growing modes ($H > 0$)and decreases
for decaying modes ($H < 0$)}.  This is of course encouraging for
the interpretation of $\ln \Omega$ as a measure of gravitational
entropy.

In one of George's preprints \cite{Ellis} he and his coauthor Reza
Tevakol mentioned that it is hard to see how to apply this
formalism to GR in full generality since the number of modes, $N$,
is infinite in field theory.  I should clear up this
misunderstanding.  We are not doing a second-quantized field
theory.  $N$ is merely the number of Planckian oscillators-- modes
in a box--which should be finite if the box under consideration is
of finite size.  To be sure, $N$ is merely the Jeans' number $\sim
\omega^3 a^3$, where $\omega$ is the oscillator frequency.  In all
our models $S \sim N$ (see especially Eq (\ref{S_ho}) below), as
is the case for electromagnetic radiation.

From this fact, we can actually calculate the entropy density of
gravitational radiation produced during inflation.  I did this
calculation quite recently and it is not to be found in Rothman
and Anninos, but it does follow closely the method outlined by
Peebles \cite{Peebles} on page 492-494. Our Lagrangian density can
be rewritten slightly as
\begin{equation}
{\cal L} = m_p^2 (\dot h^2 - h'^2/a^2)
\end{equation}
where now $\dot{ } = d/dt$.  This is a fairly generic Lagrangian
density with $m_p \dot h$ playing the role of the usual kinetic
energy $\dot\phi^2$ term. As Peebles shows, $m_p h \sim H$, the
Hubble constant, is the rms value of the field. If we assume that
$h$ represents the size of the perturbation during inflation, we
then have
 \begin{equation}
 h \sim \frac{H}{m_p} = \frac{8\pi}{3}\frac{\epsilon^2}{m_p^2},
 \end{equation}
where $\epsilon$ is the energy scale of inflation.

 Now, as
mentioned above, the entropy $S \sim N$. Define an entropy density
$\sigma \equiv N/a^3 = \omega^3$. But in our case, $\omega \sim H
\sim m_ph$ and so $\sigma \sim h^3m_p^3$. All we need to do is
scale this down to the present day to find $\sigma_o \sim
h_o^3m_p^3$, where $h_o$ is today's value of $h$. Since $\omega
\sim m_ph$, we can also write $\sigma_o \sim
(m_ph_o\omega_o)^{3/2}$. Now, $h_o$ is the strain that has just
started oscillating as its corresponding proper wavelength
$\lambda_o$ falls within the Hubble radius.  Peebles shows that
 \begin{equation}
  h_o \sim \Omega_r^{1/2}\lambda_oH_o\frac{\epsilon^2}{m_p^2}
 \end{equation}
 But $\lambda_o \sim 1/\omega_o$. Thus
 \begin{equation}
 \sigma_o \sim \Omega_r^{3/4}H_o^{3/2}\frac{\epsilon^3}{m_p^{3/2}}
 \end{equation}
 With $H_o \sim T_o^4/m_p^2$, we get
 \begin{equation}
 \sigma_o \sim \Omega_r^{3/4}T_o^3\frac{\epsilon^3}{m_p^3}.
 \label{sigma grav}
 \end{equation}
This says that the entropy of gravitational radiation produced
during inflation is supressed from that of the microwave
background by the factor $\epsilon^3/m_p^3$. Apart from the factor
 $\Omega_r^{3/4}$, which is of order unity, Eq (\ref{sigma grav}) is identical to an
equation in Brandenburger et al., obtained by much more difficult
means.

 To conclude this part of the talk, we find that the
candidate gravitational entropy function $S = \ln \Omega$ behaves
reasonably on subhorizon scales in that it increases
monotonically, and when applied to the case of gravitational
radiation during inflation produces the result that a generic
field theory should produce. Nonlinear calculations should be
carried out.

I would like now to briefly discuss two other aspects of this
project.  We did compare the behavior of $S$ with that of $C^2$
for the metric
\begin{equation}
  ds^2 = a^2(\eta) \Bigl[-(1+2\Phi(\eta,z)) d\eta^2 + (1-2\Phi(\eta,z))
         \gamma_{ij} dx^i dx^j\Bigr] ,
\label{metric_dens}
\end{equation}
where $\Phi$ is the gauge-invariant version of $h$ with solution $
\Phi = \left( \overline u_1 + \overline u_2 \eta^{-5} \right)
 e^{i(kz)}$. Specifically, we examined
\begin{equation}
C_{\alpha\beta}^{\ \ \gamma\delta} C^{\ \
\alpha\beta}_{\gamma\delta}
  = \frac{16\left(\Phi_{,zz}\right)^2}{3a^4} ,
\label{c2}
\end{equation}
where the solution for $\Phi$ is the one just given.  I bear in
mind Roger's earlier comment that he did not mean for $C^2$ to be
a measure of entropy, but with  $a \sim \eta^2$, we find that
(\ref{c2}) decreases with time and inhomogeneity, which seems a
rather strange behavior.  You might think to correct this behavior
by introducing an overall minus sign, but then the decaying modes
increase with decreasing inhomogeneity, which is no less strange.

On the other hand, we also examined the form of the Penrose
hypothesis suggested by Wainwright and collaborators
\cite{Wainwright84,Goode85,Goode92} To lowest order this is:
\begin{equation}
\frac{C_{\alpha\beta}^{\ \ \gamma\delta} C^{\ \
\alpha\beta}_{\gamma\delta}}
     {R_{\alpha\beta} R^{\alpha\beta}}
  = \frac{4a^4 \left(\Phi_{,zz}\right)^2}
         {9(\dot a^4 - a\dot a^2\ddot a + a^2 \ddot a^2)}
  = \frac{\eta^4 \left(\Phi_{,zz}\right)^2}{27} ,
\label{c2r2}
\end{equation}
which has a time dependence
\begin{eqnarray}
\frac{C_{\alpha\beta}^{\ \ \gamma\delta} C^{\ \
\alpha\beta}_{\gamma\delta}}
     {R_{\alpha\beta} R^{\alpha\beta}} \propto
 \left\{
  \begin{array}{lll}
    \eta^{4}  &,     & \qquad \mbox{for growing modes,} \\
    \eta^{-6} &,     & \qquad \mbox{for decaying modes.}
  \end{array}
  \right.
\nonumber
\end{eqnarray}
Such a time dependence is not only reasonable but is,to one's
great surprise (at least our great surprise), identical to that of
the Hamiltonian $H$ for $k = 0$ dust. The expression (\ref{c2r2})
is an approximation, so we don't know whether we are facing a
coincidence or something more profound. We haven't investigated
the matter further, but I think if there is a graduate student
interested in numerical work, it might be worth looking into.

I'd like now to turn to my recent--and so far not entirely
successful--attempt to apply this formalism to black holes. Take
expression (\ref{Omega_ho}) for the phase space of $N$ harmonic
oscillators.  Now imagine creating a black hole out of quantum
oscillators.  In this case the Hamiltonian $H$ should be the total
energy,$M$ and we should also have  $M = N\overline \epsilon$,
where $\overline\epsilon =$ average oscillator energy.  Further
the angular frequency $\omega = 2\pi\overline\epsilon= k^2$, where
$k$ is the spring constant.  Plugging all this into
(\ref{Omega_ho}) and using Stirling's approximation on the $N!$
gives identically
\begin{equation}
\Omega = e^N \mbox{and}\ S = N \label{S_ho}
\end{equation}
Again , $N = M/\overline\epsilon = M\lambda$, where $\lambda$ is
the wavelength. A priori, we expect $\lambda \approx 4M$. With
this value we find
\begin{equation}
S \approx 4M^2 = S_{BH}/(2\pi^2), \label{S_bh}
\end{equation}
where $S_{BH}$ is the canonical Bekenstein-Hawking value.

About half the people I have shown this to call it a ``discovery"
and the other half call it a ``dimensional coincidence."  I think
it is bit more than the latter.  In  his review article on string
theory and black holes, Gary Horowitz \cite{Horowitz} points out
that string entropy is given by $S_s \sim \ell_s M_s$, where
$\ell_s$ is the length of the string and $M_s$ is the string mass.
In the limit that $\ell_s$ becomes the Schwarzschild radius and
$M_s$ becomes the mass of the black hole, we get $S \sim
R_{BH}M_{BH}$. Horowitz calls this ``remarkable," yet it seems to
me to be essentially the same argument I just gave for ordinary
oscillators. The result appears to be quite general.

The main problem with this technique, as I see it, is that it
doesn't give a value for $N$, which must be put in by hand.  The
question is, if you believe $S \sim N$, can you get a better fix
on $N$?  The assumption that black-hole entropy is related to the
number of interior modes has been challenged, but in light of the
recent developments in string theory, it seems to me worth the
effort to find a more modest description along these lines.  What
I've recently been attempting to do is count interior modes of a
Schwarzschild black hole.

The idea is fairly simple.  Consider the usual scalar wave
equation on a Schwarzschild background:
\begin{eqnarray}
-(1-\frac{2M}{r})[(1-\frac{2M}{r})\psi_{,rr}+
\frac{2M}{r^2}\psi_{,r}]\nonumber \\
 +\{(1-\frac{2M}{r})[\frac{2M}{r^3}+
 \frac{\ell(\ell+1)}{r^2}]-\omega^2\} \psi & = 0
\label{scalar wv}
\end{eqnarray}
This has a well-known effective potential (see MTW, p. 868 \cite
{Misner73}), which goes to minus infinity at the origin.  But the
only thing that happens at $R = 2M$ is that $V_{eff}$ goes to
zero. Moreover, although the Legendre-functions ($P's$ and
$Q's$)that are the solutions to (\ref{scalar wv}) in the limit of
zero frequency blow up either at $2M$ or infinity (the no-hair
theorem), the $P's$ are actually regular on the interior. So the
idea is to put in a Planck-frequency cutoff $\equiv \omega_p$ and
see if one can count modes in a sensible fashion.

Now, you might think this is impossible, since (\ref{scalar wv})
has no analytic solutions on the interior, but you don't actually
need to solve the equation.  Define a new variable related to the
old by the integrating factor of the equation: $\psi =
(\frac{2M-r}{r})^{-1/2}\overline \psi$.  This eliminates the first
derivative term in the wave equation.  Recalling that $r$ is
timelike for $r < 2M$, make the ansatz  $\overline \psi \sim
e^{i\omega r}$.  Then you find
\begin{eqnarray}
 \frac{d^2\overline\psi}{dt^2}
       + [\omega^2(\frac{2M-r}{r})^2 -\frac{M^2}{r^4}
            -(\frac{2M-r}{r})  \frac{\ell(\ell+1)}{r^2}]\overline\psi
            = 0
            \label{wv eff}
\end{eqnarray}

Since $t$ is spacelike for $r < 2M$, this is just like a
flat-space wave equation,
\begin{equation}
\overline\psi'' + k_{eff}^2\overline\psi = 0,
\end{equation}
where $k_{eff}^2$ is the quantity in brackets in (\ref{wv eff}).

Now you just count modes in a box.  Following the suggestion of 't
Hooft \cite{Hooft}, the total number of radial modes beneath a
frequency $\omega$ will be,
\begin{equation}
N_{\omega} = \frac{1}{\pi}\int_{r,\ell}k_{eff} \ d\ell dr
\label{box}
\end{equation}
I claim that you can integrate this expression.  The integrals
converge at $R = 2M$ and the potential terms don't contribute. The
only thing that contributes is the cutoff frequency and the radial
cutoff ($= 2M$).  In fact, it is quite remarkable that of the six
terms in the final integral, the only non-negligible term is the
one with the correct scaling; and the result is $S =
(3/8)\pi\omega^2M^2$. Choosing the Planck frequency $\omega_p =
2\pi$ as the cutoff, the expression becomes $S = (3/16)\pi
S_{BH}$, or about .6 the canonical value. In flat space, one gets
exactly .25 $S_{BH}$.

Now, a scalar wave has only one polarization.  If one counted
gravitational perturbations instead, one would use the Zerilli or
Regge-Wheeler equation, which have different potentials.  But
since the potential terms don't contribute, one gets the same
answer per mode.  Therefore, for the two modes of gravitational
waves, the result is 1.2 $S_{BH}$, which seems too close for
comfort.

Unfortunately, there are two flies in the ointment.  The first is
that the integral over $r$ in (\ref{box}) is a timelike integral,
which makes the interpretation unclear.  More seriously, I have
ignored azimuthal modes.  To count them, the integrand in
(\ref{box}) should contain a factor $(2\ell +1)$.  However, in
that case the scaling goes wrong; i.e., one gets $S \sim w^3r^3$.

I am at this time unable to justify ignoring the azimuthal modes
and so I must let the result stand as what may well be a
dimensional fluke. Even so, it contains some interesting points in
common with earlier work of Bombelli et al. \cite{Bombelli}.  They
calculate the entropy due to a system of coupled oscillators. To
carry out the calculation they define a density matrix and perform
partial traces over an imaginary sphere that separates ``inside
oscillators" from ``outside oscillators."  In the continuum limit
they find that the entropy of the oscillators within this
imaginary sphere scales as the area.  Although the considerations
by which they obtain their results are somewhat different than
mine, they nevertheless find that the result is independent of the
potential and depends only on the cutoffs in $\omega$ and $r$ and
is proportional to the number of fields under consideration. (This
will be true whether azimuthal modes are considered or not.)
Whether this points to some deeper understanding of black hole
entropy or is merely a reflection of the fact that both
calculations are based on oscillator systems, or whether it is a
basic reflection of the dimensionful constants that govern the
problem is something that I have yet to sort out. Thank you.

\end{document}